\documentclass[twocolumn,twoside,prd,floatfix,letterpaper]{revtex4}

\usepackage{ifpdf}
\ifpdf 
  \usepackage[pdftex]{graphicx} 
\else 
  \usepackage{graphicx} 
\fi
\usepackage{amsmath}
\usepackage{amssymb}
\usepackage{fancyhdr}
\usepackage{color}
\usepackage{rotating}
\usepackage[colorlinks,hyperindex]{hyperref}
\usepackage{epigraph}
\usepackage{etoolbox}
\makeatletter
\patchcmd{\epigraph}{\@epitext{#1}}{\itshape\@epitext{#1}}{}{}
\makeatother

\def \lastDataDate {October 31, 2019}
\def \suspiciousReturns {Figures~\ref{fig:SuspiciousReturns1} and~\ref{fig:SuspiciousReturns2}}

\begin{document} 

\title{Celebrating Three Decades of Worldwide Stock Market Manipulation}
\author{Bruce Knuteson}
\noaffiliation

\begin{abstract}
  As the decade turns, we reflect on nearly thirty years of successful manipulation of the world's public equity markets.  This reflection highlights a few of the key enabling ingredients and lessons learned along the way.  A quantitative understanding of market impact and its decay, which we cover briefly, lets you move long-term market prices to your advantage at acceptable cost.  Hiding your footprints turns out to be less important than moving prices in the direction most people want them to move.  Widespread (if misplaced) trust of market prices -- buttressed by overestimates of the cost of manipulation and underestimates of the benefits to certain market participants -- makes price manipulation a particularly valuable and profitable tool.  Of the many recent stories heralding the dawn of the present golden age of misinformation, the manipulation leading to the remarkable increase in the market capitalization of the world's publicly traded companies over the past three decades is among the best.
\end{abstract}

\maketitle

\epigraph{Markets are supposed to make sense.  When you see anomalies in the market, it is probably a place where we should look further.}{--- \textup{United States Securities and Exchange Commission Chairman Jay Clayton~\cite{claytonMarketsShouldMakeSense}}}

\vspace{-0.1in}

\tableofcontents

\section{Market manipulation\label{sec:MarketManipulation}}

\begin{figure*}[ht]
  \includegraphics[width=6.5in]{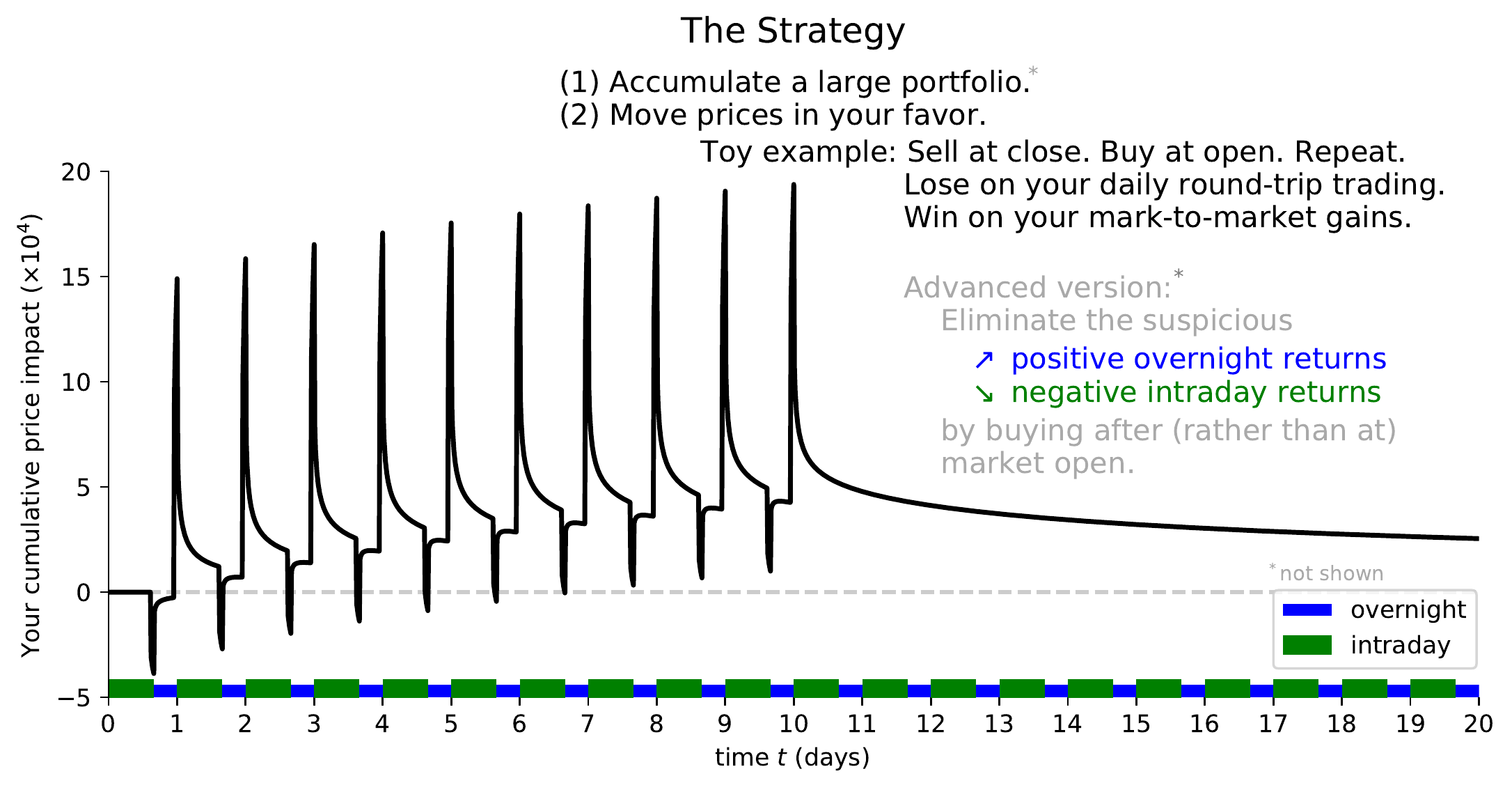}
  \caption{\label{fig:MoveThePrice}A cartoon view of the Strategy~\cite{knuteson2016information,knuteson2018wealth}.  (Real-life implementation involves more complicated trading in greater volume.)  The black curve shows the cumulative effect of your trading on price.  (A long position in a single stock is shown for simplicity.)  Prior to day 0, accumulate a large position in this stock.  On day 0 at market close ($t\approx0.66$, at the end of the first green horizontal bar), sell a small fraction of the shares you hold.  This pushes the price down by a bit less than 0.05\% (leftmost dip in black curve).  The system quickly starts to relax back to 0.  On day 1 at market open ($t=1.0$, at the end of the first blue horizontal bar), buy the same number of shares you just sold.  This pushes the price up (leftmost peak in the black curve) by 0.15\%.  The system again quickly starts to relax back to 0.  Sell the same number of shares at market close ($t\approx1.66$), buy them back at market open on day 2 ($t=2.0$), and continue this seemingly pointless daily round-trip trading.  If you were to stop after ten days (don't), the black curve from day 10 onward shows that some of your price impact sticks around for a while (as discussed in Section~\ref{sec:MarketImpact}).  Done correctly, your seemingly pointless round-trip trading creates mark-to-market gains on your large existing portfolio (accumulated before day 0) significantly exceeding the cost of this daily round-trip trading.}
\end{figure*}

Today we celebrate three decades of the Strategy~\cite{knuteson2016information,knuteson2018wealth}, a type of market manipulation employing a specific pattern of round-trip trading to create mark-to-market gains on a large existing portfolio. The Strategy, shown in cartoon form in Figure~\ref{fig:MoveThePrice}~\footnote{Practical implementation of the Strategy is significantly more complicated than the cartoon shown in Figure~\ref{fig:MoveThePrice}.  A sensible risk profile is achieved with a suitably leveraged, market-neutral portfolio covering many stocks.  Your trading, perhaps totaling in the ballpark of one percent of total market volume, will be spread throughout the day, rather than concentrated solely just before and at market open and market close as shown in Figure~\ref{fig:MoveThePrice}.  However complex the details of your trading, the important component (as far as your profits are concerned) is the expansion of your existing portfolio when the impact of your trading on the market is large and the contraction of your existing portfolio when the impact of your trading on the market is small.  (If you have lots of new money coming in from outside investors, just expand; no need to contract.)}~\footnote{The black curve in Figure~\ref{fig:MoveThePrice} shows the price change you expect in this stock due to your trading, averaging over (ignoring) stochastic variation.}, exploits a general feature of worldwide public equity markets:  early in the trading day (near market open), spreads are wide and depths are thin; late in the trading day (near market close), spreads are narrow and depths are thick~\cite{wsj2014intraday}.  An order placed near market open thus moves the price more than an equally sized order placed near market close~\cite{cont2014price}~\footnote{This intraday predictability is a reasonable, perfectly understandable, and not-intrinsically-problematic feature of public stock markets.  Market makers, not wishing to be on the wrong side of overnight news they may have missed, make wider markets early in the trading day.  The market, viewed as an information aggregator, respects the information content of orders placed near the start of the trading day more than the information content of orders placed later in the trading day.}~\footnote{Considering a snapshot in time near market open (and speaking loosely to help the non-expert build intuition), ``spreads are wide and depths are thin'' means there are few other orders near the fair market price.  Considering a snapshot in time near market close, ``spreads are narrow and depths are thick'' means there are many other orders near the fair market price.  You can expect your impact to be greater when you are one of a few than when you are one of many.}.

The Strategy is to accumulate a large portfolio and then systematically and repeatedly expand it a bit near market open (when spreads are wide and depths are thin) and contract it a bit near market close (when spreads are narrow and depths are thick)~\footnote{That such market manipulation is possible is not in question.  The cost of each round-trip trade depends on how much you trade, but does not depend on the size of your existing portfolio.  Your mark-to-market gains are proportional to the size of your existing portfolio.  Your mark-to-market gains will therefore exceed the cost of your round-trip trading as long as your existing portfolio is sufficiently large.  The practical threshold for ``sufficiently large'' is in question, but for the world's stock markets, roughly one billion dollars of capital (suitably leveraged, and used to form a market-neutral equity portfolio covering many stocks) appears sufficient~\cite{knuteson2016information}.}.  Think of your portfolio breathing, expanding its lungs near market open and contracting its lungs near market close.  On each breath, the number of shares in equals the number of shares out, but in units of dollars, your lungs expand just a bit more when you inhale than when you exhale, leaving the value of your portfolio (in units of dollars) a bit bigger at the end of each breath.

If you are a scientist or engineer, think of the market as a simple mechanical system.  Placing an order near market open perturbs this system.  The system then relaxes (with part of this relaxation happening quickly and some of it happening slowly, as discussed further in Section~\ref{sec:MarketImpact}).  Placing a similarly-sized order near market close perturbs the system less.  The response of the system to your perturbations (and the way it subsequently relaxes) can be modeled and understood the same way you would model and understand any other system:  systematically perturb it and note what happens.  There is no magic here.

Each of your daily round trips, considered in isolation, is a money-losing effort.  Properly and repeatedly done, however, your daily round trips push prices in your favor, causing mark-to-market gains on your large existing portfolio.  The Strategy's pattern of round-trip trading -- expanding your existing portfolio when your trading moves prices more, contracting it when your trading moves prices less, losing money on your round-trip trading, and posting mark-to-market gains on your large existing portfolio due to the price impact of your trades -- is market manipulation under any reasonable definition.

\begin{figure*}[p]
\includegraphics[width=7in]{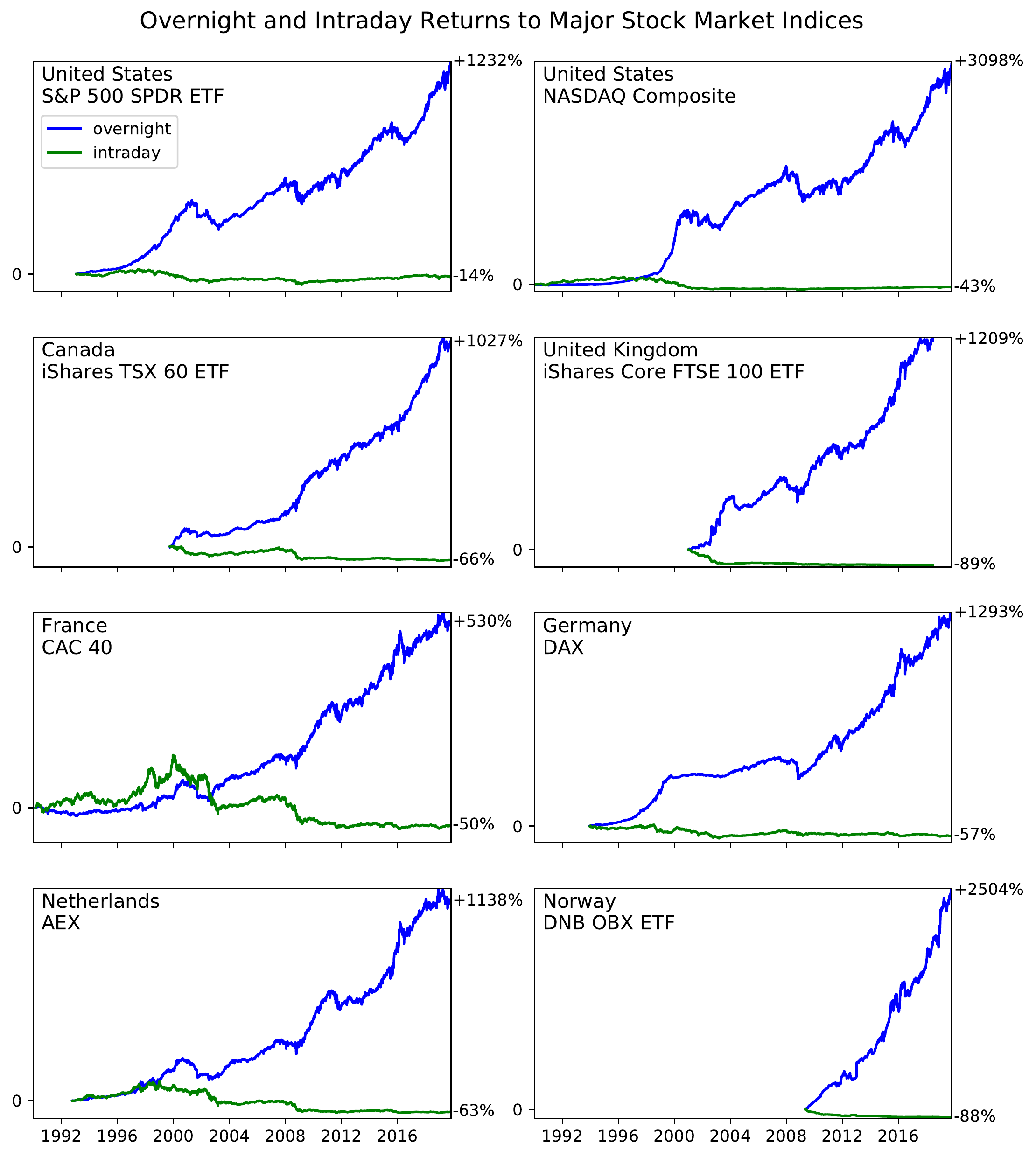}
\caption{\label{fig:SuspiciousReturns1}Cumulative overnight (blue curve) and intraday (green curve) returns to eight major stock market indices over three decades.  The overnight (blue) curve cumulates returns from market close to the next day's market open.  The intraday (green) curve cumulates returns from market open to market close.  The horizontal axis of each plot extends from  January 1, 1990 to \lastDataDate.  The (linear) vertical scale in each plot extends from a return of -100\% (bottom of plot) through 0 (explicitly marked, at left) to the largest cumulative overnight return achieved (top of plot).  On each plot, the cumulative overnight and intraday returns on \lastDataDate\ (or the last date available) are explicitly marked, at right.  Several curves start on the first day for which data are available:  S\&P 500 (1993-01-29), TSX 60 (1999-10-04), FTSE 100 (2001-01-02 to 2018-06-20), CAC 40 (1990-03-01), DAX (1993-12-14), AEX (1992-10-12), and DNB OBX (2009-05-08).  The code used to make this figure is available at Ref.~\cite{thisArticleWebpage}.  Data are publicly available from Yahoo!~Finance.}
\end{figure*}

\begin{figure*}[p]
\includegraphics[width=7in]{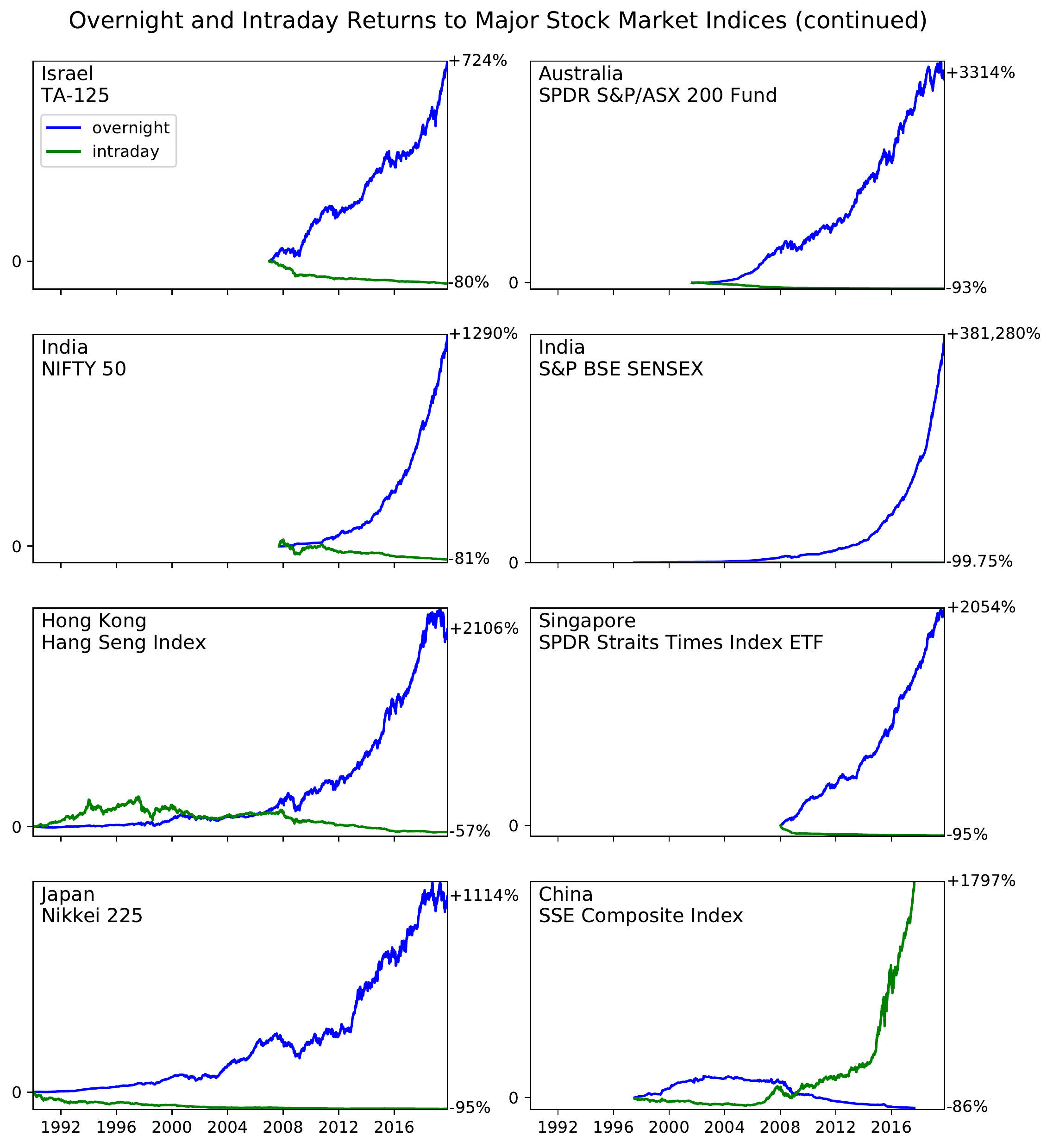}
\caption{\label{fig:SuspiciousReturns2}Cumulative overnight and intraday returns to eight more major stock market indices, prepared in the same manner as Figure~\ref{fig:SuspiciousReturns1}.  Several curves start on the first day for which data are available: TA-125 (2007-01-08), ASX 200 (2001-08-27), NIFTY 50 (2007-09-17), SENSEX (1997-07-01), Straits Times (2008-01-10), and SSE (1997-07-02 to 2017-08-25).  SENSEX prices from the Bombay Stock Exchange (available from 2009 onwards)~\cite{bseIndiaData} match those from Yahoo!~Finance used for this plot.  China's return pattern can be understood in terms of China's ``T+1'' trading rule (which makes the ``expand your longs in the morning and contract them in the afternoon'' half of the Strategy explicitly-and-easily-enforceably illegal~\cite{qiao2019china}) and China's ban on short selling before 2010 (making the ``expand your shorts in the morning and contract them in the afternoon'' half of the Strategy impossible before 2010)~\cite{chang2014short}.}
\end{figure*}

Note that the cartoon form of the Strategy drawn in Figure~\ref{fig:MoveThePrice} creates a suspicious return pattern.  The intraday return (from market open to market close) corresponds to the negative return from peak to dip during each intraday period (marked with a green horizontal bar along the x-axis).  The overnight return (from market close to the next day's market open) corresponds to the positive return from the bottom of each dip to the top of the peak the next day during each overnight period (marked with a blue horizontal bar along the x-axis).  Stitching these together day after day, this cartoon view of the Strategy produces a sequence of positive overnight returns and negative intraday returns similar to those seen in the world's major stock market indices over the past three decades, as shown in \suspiciousReturns~\footnote{Most of the plots in \suspiciousReturns\ actually underestimate the true divergence between overnight and intraday returns over the past three decades.  Many of the plots of indices (such as the NASDAQ Composite index) do not include dividends with reinvestment, the inclusion of which leaves the green (intraday) curve unchanged and further increases the height of the blue (overnight) curve.  (The most recent mainstream news article we are aware of covering the first plot in Figure~\ref{fig:SuspiciousReturns1}~\cite{sommer2018night} excludes dividends, thereby understating overnight returns by nearly a factor of two.)  As a separate matter, correcting stale opening prices in index constituents (see Table 3 of Ref.~\cite{lachance2015night}) further increases the divergence between the overnight and intraday curves in the plots of indices shown in \suspiciousReturns, which do not include this correction.}.

The first plot in Figure~\ref{fig:SuspiciousReturns1} appeared in Ref.~\cite{cooper2008return} twelve years ago.  Most of the remaining content of \suspiciousReturns\ appeared in Ref.~\cite{lachance2015night} over four years ago.

The obvious, mechanical explanation of the highly suspicious return patterns shown in \suspiciousReturns\ is someone trading in a way that pushes prices up before or at market open, thus causing the blue curve, and then trading in a way that pushes prices down between market open (not including market open) and market close (including market close), thus causing the green curve.  The consistency with which this is done points to the actions of a few quantitative trading firms rather than the uncoordinated, manual trading of millions of people.  Ref.~\cite{cooper2008return} concluded as much twelve years ago, ending with the paragraph:

\begin{quote}
Hopefully, future extensions of our results will help explain further the sources of the day and night effect. Potential explanations may come from an examination of the effects of the growing and widespread practice of algorithmic trading by hedge funds and other financial institutions; perhaps price pressure effects from algorithm generated trading may account for some of the observed price patterns we document.
\end{quote}

The existence of the Strategy explains how such a firm (with the qualities described in Ref.~\cite{knuteson2018wealth}) can benefit from this seemingly pointless and costly price pushing.  The literature currently contains zero plausible alternative explanations for these highly suspicious return patterns in the world's major stock market indices~\footnote{The attempted explanation we hear most frequently is that ``company news'' (particularly quarterly earnings) is often announced overnight and over the past three decades this news has generally been good~\cite{mccrum2018SomeoneIsWrong}.  Section~4.1 of Ref.~\cite{cooper2008return} dispensed with this attempted explanation twelve years ago:  removing the days corresponding to company earnings announcements does not change the overnight/intraday split shown in the first plot in Figure~\ref{fig:SuspiciousReturns1} in the slightest.  Separately, no analysis whatsoever is required to see that the release of company news overnight does not explain the consistently negative intraday returns shown in \suspiciousReturns.}~\footnote{We would obviously be very happy to find that the cause of the highly suspicious return patterns in \suspiciousReturns\ is innocuous.  We consider this unlikely, in part because it would be the first time in the history of financial markets that highly suspicious return patterns turned out to be fine.}.

The last plot in Figure~\ref{fig:SuspiciousReturns2} is a fun variation on the general theme in \suspiciousReturns.  China is unique in having a ``T+1'' trading rule that prohibits your buying a share of a company and then selling it later the same day~\cite{qiao2019china}, making the ``expand your long positions in the morning and contract them in the afternoon'' half of the Strategy explicitly-and-easily-enforceably illegal~\footnote{The Strategy typically involves systematically expanding and contracting a market-neutral equity portfolio consisting of both long and short positions.  In some cases, including when regulations differ in their treatment of your long and short positions, it may be convenient to talk about the expansion and contraction of your long positions as one ``half'' of the Strategy, and the expansion and contraction of your short positions as the other ``half'' of the Strategy.}.  China, which legalized short selling in 2010~\cite{chang2014short}, is totally fine with your shorting a share of a company and then buying it back later the same day, making the ``expand your short positions in the morning and contract them in the afternoon'' half of the Strategy legal (or, equivalently, not-easily-enforceably illegal).  As expected, the pattern of overnight and intraday returns in China in the last plot in Figure~\ref{fig:SuspiciousReturns2}~\footnote{We thank Kenan Qiao and Lammertjan Dam, the authors of Ref.~\cite{qiao2019china}, for helpful comments and analysis provided in private correspondence related to the last plot in Figure~\ref{fig:SuspiciousReturns2}.} is consistent with firms executing the not-explicitly-and-easily-enforceably illegal half of the Strategy from 2010 onward.

An earlier caution~\cite{knuteson2018wealth} bears repeating: implementing the Strategy in a manner creating the highly suspicious return patterns shown in \suspiciousReturns\ is unnecessary.  In Figure~\ref{fig:MoveThePrice}, move your morning buy from just before or at market open to just after market open so you do not leave the glaringly obviously problematic return patterns shown in \suspiciousReturns~\footnote{Doing less of your morning expansion before and at market open in the United States after Ref.~\cite{cooper2008return} pointed out the first plot in Figure~\ref{fig:SuspiciousReturns1} has reduced the highly suspicious divergence between overnight and intraday returns in the United States from 2008 onward (a fact more obvious when the top row of Figure~\ref{fig:SuspiciousReturns1} is plotted starting in 2008).  This shift in your morning trading, combined with your regulator's inability to see how you are trading without your explicit assistance~\cite{claytonConsolidatedAuditTrail}, has facilitated your continued, unhampered use of the Strategy in the United States.}.  Trading in a way that leaves \suspiciousReturns\ is roughly equivalent to committing crimes and leaving smoking guns right in front of police stations in fourteen separate jurisdictions, with your fingerprints all over the guns.  In India, where you can take the Strategy to a completely different level~\footnote{The Bombay Stock Exchange provides SENSEX prices from 2009-01-01 onward~\cite{bseIndiaData}.  These prices agree (with a few immaterial exceptions) with the prices provided by Yahoo!~Finance during this time.}, you have left a smoking bazooka.  Don't do this.  There are plenty of ways to implement the Strategy without leaving ridiculous price patterns for anyone to see in data publicly available from Yahoo!~Finance.

Fortunately, human nature being what it is, most are happy to ignore smoking guns if doing so increases the balance in their retirement accounts.  The disconnect between this willful blindness and the quote starting this article is stark.

\section{Market impact\label{sec:MarketImpact}}

Market manipulation is similar in nature to other misinformation campaigns in that effective implementation is aided by an accurate understanding of the costs and benefits of actions available to you.  With the Strategy, accurately predicting both costs (losses on your daily round-trip trading) and benefits (mark-to-market gains resulting from you pushing prices in your favor) requires understanding how much the market changes when you perturb it by submitting an order.  The impact your order has on the market is called ``market impact''~\footnote{The phrase ``initial impact'' (or ``instantaneous impact'') refers to the immediate impact of your order (at $t=0$, the time your order is placed, before any subsequent relaxation).  The phrase ``impact decay'' refers to the market subsequently relaxing (during times $t>0$).    We use the phrases ``market impact'' and ``price impact'' (or in certain places, to avoid ambiguity, ``market impact and decay'') to refer to a full function of time (defined for $t \ge 0$), encompassing both ``initial impact'' (at $t=0$) and ``impact decay'' (during subsequent times $t>0$).}.

When celebrating a technical achievement, it is permissible to indulge in a few details as long as they are kept short.  We permit ourselves a one-paragraph summary of the existing literature on market impact.

\begin{figure}[htb]
\includegraphics[width=3.5in]{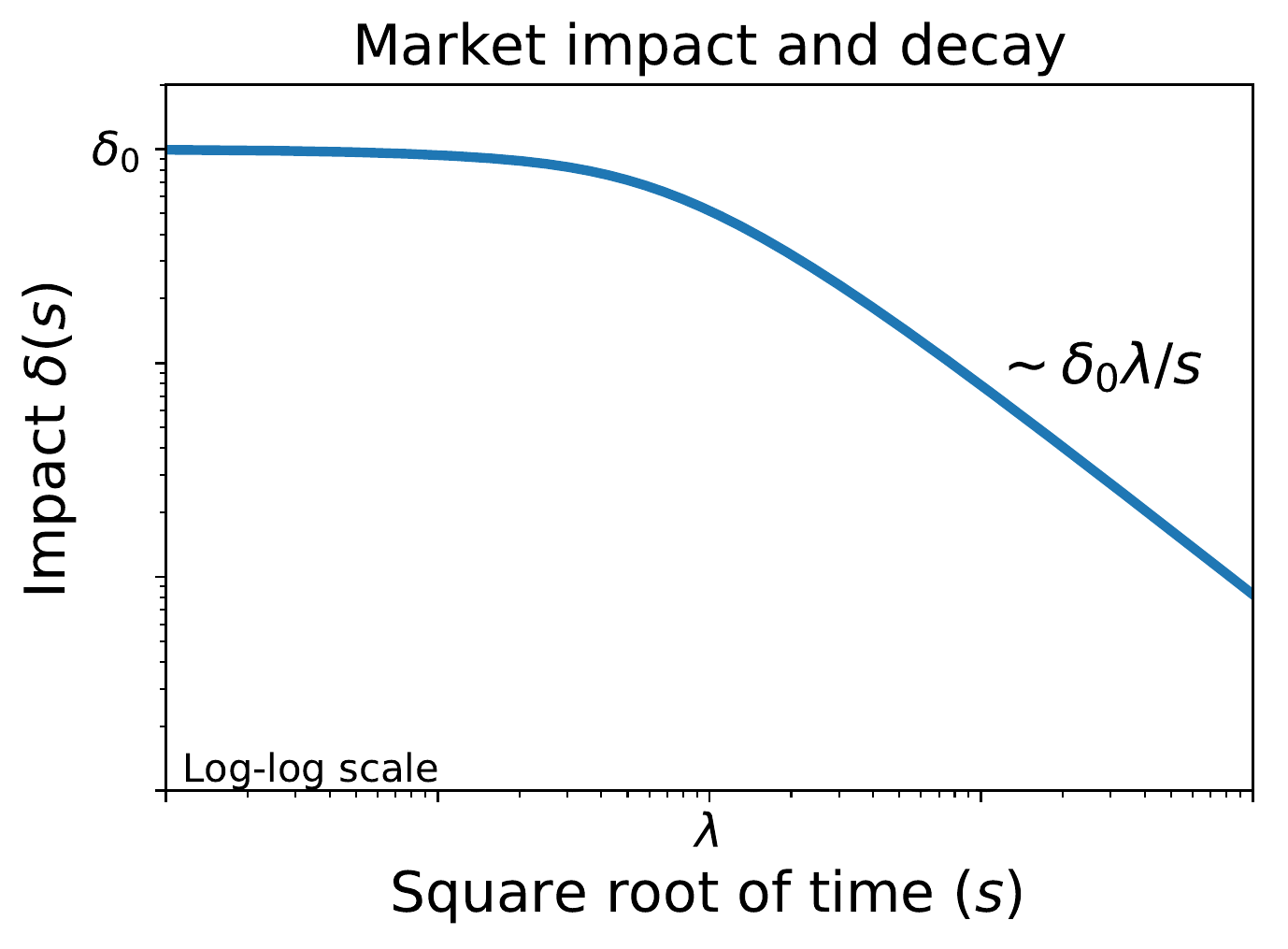}
\caption{\label{fig:ImpactAndDecay}The blue curve shows the price impact $\delta(s)$ of a single order (vertical axis, logarithmic scale) as a function of $s$, the square root of the time elapsed after order placement (horizontal axis, logarithmic scale).  The intercept $\delta(0)$ is, tautologically, your best estimate of your order's initial price impact, $\delta_0$.  Refs.~\cite{gatheral2010no,donier2015fully,benzaquen2018market} imply this initial market impact asymptotically decays as $s^{-1}$.  These two boundary conditions constrain $\delta(s)$ to something like the blue curve shown.  The curve must have a ``knee'' (power law cutoff), the abscissa of which we denote by $\lambda$.  These facts together imply $\delta(s) \xrightarrow{s \gg \lambda} \delta_0 \lambda / s$, up to a multiplicative constant of order unity.}
\end{figure}

Considering the limit order book of an active market, define a ``fair market price'' that probably lies somewhere between the best bid and best offer.  Imagine placing a single order.  Given your knowledge of the current state of the limit order book, let $\delta_0$ denote the fractional change in fair market price you expect placing your order to have immediately upon placement~\footnote{For example, if the ``fair market price'' before your order is \$10.00 and you expect the ``fair market price'' after you place your order to be \$10.002, then $\delta_0=0.0002$ ($=(10.002 - 10.00) / 10.00$).}.  Letting $t$ denote the time elapsed after the placing of your order and introducing the parameter $s(t)=\sqrt{t}$ for the sole purpose of using fewer square root signs, Refs.~\cite{gatheral2010no,donier2015fully,benzaquen2018market} imply your price impact asymptotically decays as $s^{-1}$.  That is, your best estimate of your order's initial price impact is (tautologically) $\delta_0$, and the existing literature claims this impact decays as $s^{-1}$ for sufficiently large $s$.  Graphing this on a log-log scale~\footnote{In linear scale, the curve in Figure~\ref{fig:ImpactAndDecay} approximates one of the ten peak-and-decays (starting at market open) shown in Figure~\ref{fig:MoveThePrice}.}, as in Figure~\ref{fig:ImpactAndDecay}, the only other point of interest is the knee, which occurs at $s\approx\lambda$~\footnote{As far as we know, the correct expression for $\lambda$ in terms of other relevant quantities has not yet made its way into the public domain.}.

For the manipulator, the takeaways from the preceding one-paragraph literature summary and Figure~\ref{fig:ImpactAndDecay} (in light of the predictable variation in spreads and depths over the course of the trading day noted in Section~\ref{sec:MarketManipulation}) are that (i) you can submit two similarly-sized orders with significantly and predictably different price impact by carefully choosing when you place them (through $\delta_0$, and perhaps also through $\lambda$), and (ii) some of your price impact sticks around a long time, allowing persistent price pushes to accumulate materially in your favor.  These are the features of market impact that make the Strategy possible in practice.

\section{Human manipulation\label{sec:HumanManipulation}}

All subjects of celebration have a supporting cast.  We wish to acknowledge two in particular.

First, the academic peer review process has supported continued use of the Strategy with both false positives and false negatives.  Accepting studies of questionable accuracy, the peer review system has provided trading strategies quants have used to construct portfolios.  Since what matters most is accumulating a large portfolio and then trading at the margin to move prices in your favor (as shown in Figure~\ref{fig:MoveThePrice}), the (in)accuracy of the majority of these studies is not a problem.  Constructing signals from the same pool of academic work has helped align the portfolios of multiple firms, facilitating constructive interference in their use of the Strategy.  On the flip side, failure to accept Refs.~\cite{cooper2008return,lachance2015night} into the peer-reviewed literature has contributed to a surprising but very fortunate general ignorance regarding the existence of the highly suspicious return patterns shown in \suspiciousReturns.  The academic peer review process appears to have failed precisely where it was needed most, and exactly as we might have hoped and expected given its design.

Second, and more importantly, the successful use of the Strategy over three decades has required the careful and thoughtful manipulation of individual people.  This must continue.

The incentives described in Ref.~\cite{knuteson2018wealth} keep those in your surrounding ecosystem in check and provide near immunity to prosecution as long as your use of the Strategy is successful, but nothing good lasts forever.  Avoiding subsequent prosecution requires your not being among the topmost decision makers in your firm who are on record as being aware of your use of the Strategy~\footnote{Not to belabor the point, but use of the Strategy is illegal.  Losing money on daily round-trip trading in a way that moves prices to your benefit is market manipulation.  Trading in this manner with money from outside investors without disclosing it is fraud.}.

If you are already on record as being aware of your previous or ongoing use of the Strategy, you must communicate this fact as a potential problem (clearly and on the record, multiple times) to the person directly above you, transferring as much of your personal liability as possible to him~\footnote{Nearly everyone involved here is male.}.

If you are among the top of the chain of command, you must ensure none of the people below you put you clearly on record as being aware of your material use of the Strategy.  Maintaining plausible deniability requires attentive and dexterous manipulation of the people immediately around you.

Any potential problem is apt to develop along a predictable path.  An underling with overly rigid ethical sensibilities understanding the materiality of your use of the Strategy will begin agitating to his peers and manager.  Earlier restriction and careful culling of your direct reports will ensure this quant's manager is not you.  Observing such agitation, you must eliminate this employee as quickly as possible.  Your action, which must be framed to appear to others as well-reasoned and thoughtfully justified, and in which you will need to involve others (including human resources and perhaps other colleagues) to diffuse your liability, must be brutal in its speed and effectiveness.  The time elapsed should be days, not weeks, and certainly not months.  Every passing day potentially adds to the document trail you must ensure you are not on.

If the problematic employee agitates to colleagues in other departments (including legal and compliance) or regulators, they will come to you, and you will be able to satisfy their concern with some appropriate version of there being nothing to see here.  Penetrating follow-up questions, the formulation of which requires at least a basic level of knowledge, will not be forthcoming.  Such interactions are problematic in the long term to the extent they further entangle you in an incriminating document trail.

\section{Human impact\label{sec:HumanImpact}}

Three decades of worldwide stock market manipulation is quite an accomplishment, but it is the wider human impact explained in Ref.~\cite{knuteson2018wealth} that makes this achievement particularly special.

The tens of trillions of dollars your use of the Strategy has created out of thin air have mostly gone to the already-wealthy: company executives and existing shareholders benefitting directly from rising stock prices; owners of private companies and other assets, including real estate, whose values tend to rise and fall with the stock market; and those in the financial industry and elsewhere with opportunities to ``privatize the gains and socialize the losses,'' as those in the business of doing so like to say.  These gains to capital over the last three decades have contributed directly and significantly to the current level of wealth inequality in the United States and elsewhere~\cite{kuhn2017income,piketty2014capital}.  As a general matter, widespread mispricing leads to misallocation of capital and human effort, and widespread inequality negatively affects our social structure and the perceived social contract.

The fact that three decades of small price nudges by so few can have such far-reaching consequences in so many areas of human life is truly marvelous.

\section{Be careful\label{sec:BeCareful}}

There is little more we can do here.  You are now finishing the last of a trilogy of articles~\cite{knuteson2016information,knuteson2018wealth}, written over as many years, of gradually increasing scope.  We have repeatedly brought this matter to the attention of relevant regulators, both domestic and foreign, and to hundreds of journalists, academics, and other professionals.  None have offered an alternative plausible explanation for the highly suspicious return patterns shown in \suspiciousReturns.  None have offered evidence disfavoring the explanation provided in this article.  These efforts, spread over half a decade, have led to a grand total of one-third of an article written by somebody else~\cite{bloomberg2018palantir}.

Our younger selves would have been confident that somebody somewhere was working to fix this~\footnote{The correct fix is to understand the cause of the suspicious return patterns in \suspiciousReturns\ and to communicate this to the public clearly and transparently, trusting our free markets to properly incorporate this new information.  The correct fix is not to pour significant public resources into hiding the problem, hoping nobody notices.  The financial system is largely built on trust, and trust can be a fragile thing.}, and that there will always be at least a few willing to work against their immediate self-interest to protect others from easily avoidable harm.  Older and wiser, we are no longer so sure.

Perhaps somebody else will step up.  Perhaps not.  Today we celebrate nearly three decades of worldwide stock market manipulation.  Tomorrow circumstances may change.  At some point hundreds of millions of people will realize they have been had.  Their anger will be fully justified.  You will not want it directed at you.

\bibliography{mm}

\begin{thebibliography}{20}
\expandafter\ifx\csname natexlab\endcsname\relax\def\natexlab#1{#1}\fi
\expandafter\ifx\csname bibnamefont\endcsname\relax
  \def\bibnamefont#1{#1}\fi
\expandafter\ifx\csname bibfnamefont\endcsname\relax
  \def\bibfnamefont#1{#1}\fi
\expandafter\ifx\csname citenamefont\endcsname\relax
  \def\citenamefont#1{#1}\fi
\expandafter\ifx\csname url\endcsname\relax
  \def\url#1{\texttt{#1}}\fi
\expandafter\ifx\csname urlprefix\endcsname\relax\def\urlprefix{URL }\fi
\providecommand{\bibinfo}[2]{#2}
\providecommand{\eprint}[2][]{\url{#2}}

\bibitem[{cla()}]{claytonMarketsShouldMakeSense}
\bibinfo{note}{{The Securities and Exchange Commission: Priorities Going
  Forward}. September 5, 2017, New York, NY.
  \url{https://www.youtube.com/watch?v=ln0hlt2IywY&t=2951}}.

\bibitem[{\citenamefont{Knuteson}(2016)}]{knuteson2016information}
\bibinfo{author}{\bibfnamefont{B.}~\bibnamefont{Knuteson}},
  \emph{\bibinfo{title}{{Information, Impact, Ignorance, Illegality, Investing,
  and Inequality}}} (\bibinfo{year}{2016}),
  \urlprefix\url{https://arxiv.org/abs/1612.06855}.

\bibitem[{\citenamefont{Knuteson}(2018)}]{knuteson2018wealth}
\bibinfo{author}{\bibfnamefont{B.}~\bibnamefont{Knuteson}},
  \emph{\bibinfo{title}{{How to Increase Global Wealth Inequality for Fun and
  Profit}}} (\bibinfo{year}{2018}),
  \urlprefix\url{https://ssrn.com/abstract=3282845}.

\bibitem[{\citenamefont{Strumpf and Driebusch}()}]{wsj2014intraday}
\bibinfo{author}{\bibfnamefont{D.}~\bibnamefont{Strumpf}} \bibnamefont{and}
  \bibinfo{author}{\bibfnamefont{C.}~\bibnamefont{Driebusch}},
  \emph{\bibinfo{title}{{Why Morning Is the Worst Time to Trade Stocks}}},
  \bibinfo{howpublished}{\textit{The Wall Street Journal}, September 14, 2015},
  \urlprefix\url{http://www.wsj.com/articles/early-birds-suffer-in-market-1442273794}.

\bibitem[{\citenamefont{Cont et~al.}(2014)\citenamefont{Cont, Kukanov, and
  Stoikov}}]{cont2014price}
\bibinfo{author}{\bibfnamefont{R.}~\bibnamefont{Cont}},
  \bibinfo{author}{\bibfnamefont{A.}~\bibnamefont{Kukanov}}, \bibnamefont{and}
  \bibinfo{author}{\bibfnamefont{S.}~\bibnamefont{Stoikov}},
  \bibinfo{journal}{Journal of Financial Econometrics}
  \textbf{\bibinfo{volume}{12}}, \bibinfo{pages}{47} (\bibinfo{year}{2014}),
  \urlprefix\url{https://arxiv.org/abs/1011.6402}.

\bibitem[{thi()}]{thisArticleWebpage}
\bibinfo{note}{\url{https://bruceknuteson.github.io/spy-day-and-night}}.

\bibitem[{bse()}]{bseIndiaData}
\bibinfo{note}{\url{https://www.bseindia.com/indices/IndexArchiveData.html}}.

\bibitem[{\citenamefont{Qiao and Dam}(2019)}]{qiao2019china}
\bibinfo{author}{\bibfnamefont{K.}~\bibnamefont{Qiao}} \bibnamefont{and}
  \bibinfo{author}{\bibfnamefont{L.}~\bibnamefont{Dam}},
  \emph{\bibinfo{title}{{The Overnight Return Puzzle and the ``T+1" Trading
  Rule in Chinese Stock Markets}}} (\bibinfo{year}{2019}),
  \urlprefix\url{https://ssrn.com/abstract=3418356}.

\bibitem[{\citenamefont{Chang et~al.}(2014)\citenamefont{Chang, Luo, and
  Ren}}]{chang2014short}
\bibinfo{author}{\bibfnamefont{E.~C.} \bibnamefont{Chang}},
  \bibinfo{author}{\bibfnamefont{Y.}~\bibnamefont{Luo}}, \bibnamefont{and}
  \bibinfo{author}{\bibfnamefont{J.}~\bibnamefont{Ren}},
  \bibinfo{journal}{Journal of Banking \& Finance}
  \textbf{\bibinfo{volume}{48}}, \bibinfo{pages}{411} (\bibinfo{year}{2014}),
  \urlprefix\url{https://ssrn.com/abstract=2135067}.

\bibitem[{\citenamefont{Cooper et~al.}(2008)\citenamefont{Cooper, Cliff, and
  Gulen}}]{cooper2008return}
\bibinfo{author}{\bibfnamefont{M.~J.} \bibnamefont{Cooper}},
  \bibinfo{author}{\bibfnamefont{M.~T.} \bibnamefont{Cliff}}, \bibnamefont{and}
  \bibinfo{author}{\bibfnamefont{H.}~\bibnamefont{Gulen}},
  \emph{\bibinfo{title}{{Return Differences Between Trading and Non-trading
  Hours: Like Night and Day}}} (\bibinfo{year}{2008}),
  \urlprefix\url{http://ssrn.com/abstract=1004081}.

\bibitem[{\citenamefont{Lachance}(2015)}]{lachance2015night}
\bibinfo{author}{\bibfnamefont{M.-E.} \bibnamefont{Lachance}},
  \emph{\bibinfo{title}{{Night Trading: Lower Risk But Higher Returns?}}}
  (\bibinfo{year}{2015}), \urlprefix\url{https://ssrn.com/abstract=2633476}.

\bibitem[{\citenamefont{Gatheral}(2010)}]{gatheral2010no}
\bibinfo{author}{\bibfnamefont{J.}~\bibnamefont{Gatheral}},
  \bibinfo{journal}{Quantitative Finance} \textbf{\bibinfo{volume}{10}},
  \bibinfo{pages}{749} (\bibinfo{year}{2010}),
  \urlprefix\url{https://ssrn.com/abstract=1292353}.

\bibitem[{\citenamefont{Donier et~al.}(2015)\citenamefont{Donier, Bonart,
  Mastromatteo, and Bouchaud}}]{donier2015fully}
\bibinfo{author}{\bibfnamefont{J.}~\bibnamefont{Donier}},
  \bibinfo{author}{\bibfnamefont{J.}~\bibnamefont{Bonart}},
  \bibinfo{author}{\bibfnamefont{I.}~\bibnamefont{Mastromatteo}},
  \bibnamefont{and} \bibinfo{author}{\bibfnamefont{J.-P.}
  \bibnamefont{Bouchaud}}, \bibinfo{journal}{Quantitative Finance}
  \textbf{\bibinfo{volume}{15}}, \bibinfo{pages}{1109} (\bibinfo{year}{2015}),
  \urlprefix\url{http://arxiv.org/abs/1412.0141}.

\bibitem[{\citenamefont{Benzaquen and Bouchaud}(2018)}]{benzaquen2018market}
\bibinfo{author}{\bibfnamefont{M.}~\bibnamefont{Benzaquen}} \bibnamefont{and}
  \bibinfo{author}{\bibfnamefont{J.-P.} \bibnamefont{Bouchaud}},
  \bibinfo{journal}{Quantitative Finance} \textbf{\bibinfo{volume}{18}},
  \bibinfo{pages}{1781} (\bibinfo{year}{2018}),
  \urlprefix\url{https://arxiv.org/abs/1710.03734}.

\bibitem[{\citenamefont{Kuhn et~al.}(2017)\citenamefont{Kuhn, Schularick, and
  Steins}}]{kuhn2017income}
\bibinfo{author}{\bibfnamefont{M.}~\bibnamefont{Kuhn}},
  \bibinfo{author}{\bibfnamefont{M.}~\bibnamefont{Schularick}},
  \bibnamefont{and} \bibinfo{author}{\bibfnamefont{U.}~\bibnamefont{Steins}}
  (\bibinfo{year}{2017}), \bibinfo{note}{{CEPR Discussion Paper No. DP12218.}},
  \urlprefix\url{https://ssrn.com/abstract=3018472}.

\bibitem[{\citenamefont{Piketty}(2014)}]{piketty2014capital}
\bibinfo{author}{\bibfnamefont{T.}~\bibnamefont{Piketty}},
  \emph{\bibinfo{title}{Capital in the Twenty-First Century}}
  (\bibinfo{publisher}{Harvard University Press}, \bibinfo{year}{2014}), ISBN
  \bibinfo{isbn}{9780674369559},
  \urlprefix\url{https://books.google.com/books?id=J222AgAAQBAJ}.

\bibitem[{\citenamefont{Levin}()}]{bloomberg2018palantir}
\bibinfo{author}{\bibfnamefont{M.}~\bibnamefont{Levin}},
  \emph{\bibinfo{title}{{Nobody Knows What Palantir Is Worth}}},
  \bibinfo{howpublished}{\textit{Bloomberg Opinion}, November 15, 2018},
  \urlprefix\url{https://www.bloomberg.com/opinion/articles/2018-11-15/nobody-knows-what-palantir-is-worth}.

\bibitem[{\citenamefont{Sommer}()}]{sommer2018night}
\bibinfo{author}{\bibfnamefont{J.}~\bibnamefont{Sommer}},
  \emph{\bibinfo{title}{{The Stock Market Works by Day, But It Loves the
  Night}}}, \bibinfo{howpublished}{\textit{The New York Times}, February 2,
  2018},
  \urlprefix\url{https://www.nytimes.com/2018/02/02/your-money/stock-market-after-hours-trading.html}.

\bibitem[{\citenamefont{McCrum}()}]{mccrum2018SomeoneIsWrong}
\bibinfo{author}{\bibfnamefont{D.}~\bibnamefont{McCrum}},
  \emph{\bibinfo{title}{{Someone is Wrong on the Internet, Day Versus Night
  Edition}}}, \bibinfo{howpublished}{\textit{Financial Times}, February 6,
  2018},
  \urlprefix\url{https://ftalphaville.ft.com/2018/02/06/2198427/someone-is-wrong-on-the-internet-day-versus-night-edition/}.

\bibitem[{\citenamefont{Clayton}()}]{claytonConsolidatedAuditTrail}
\bibinfo{author}{\bibfnamefont{J.}~\bibnamefont{Clayton}},
  \emph{\bibinfo{title}{{Statement on Status of the Consolidated Audit
  Trail}}}, \bibinfo{howpublished}{September 9, 2019},
  \urlprefix\url{https://www.sec.gov/news/public-statement/statement-status-consolidated-audit-trail}.

\end{thebibliography}

\end{document}